# CRITICAL BEHAVIOR OF A FOUR-POINT SCHWINGER-DYSON EQUATION

B. Holdom[*]    and G. Triantaphyllou [†]

Department of Physics, University of Toronto

Toronto, ON M5S 1A7 Canada

May 22, 1995

## Abstract

We study the Schwinger-Dyson equation associated with a chirality-changing fermion 4-point function in a strongly-coupled $U(1)$ gauge theory. After making appropriate simplifications, we solve the equation numerically via a relaxation method. Our analysis provides an estimate of the critical coupling and it gives some indication as to the general momentum dependence of the 4-point function.

---

[*]e-mail:holdom@utcc.utoronto.ca

[†]e-mail:george@medb.physics.utoronto.ca

# 1 Introduction

In this paper we wish to study the dynamical generation of momentum dependent fermion 4-point functions within strongly interacting gauge theories. These will be purely nonperturbative quantities associated with breakdown of chiral symmetries, and would for example imply the existence of the corresponding 4-fermion condensates. Our analysis is based on the Schwinger-Dyson (SD) formalism. The resulting equations are analytically intractable and thus we seek numerical solutions.

We have two main goals in this study. One is to estimate the "critical coupling" necessary for the formation of the 4-fermion condensates. We would like to study this in the context of a situation where 4-fermion condensates develop on scales higher than the scale of mass formation. This would make it consistent to treat the 4-fermion condensate problem in isolation, independently of the mass generation problem. (This is not the case for QCD, but even in that case it is useful to at least consider the existence of the effects we study here). In extensions of the standard model the hierarchical symmetry breaking pattern we envisage can be a natural consequence of the fact that a 4-fermion condensate may break fewer gauged symmetries than the 2-fermion condensate.

For example, in [1] it is shown how the top mass may be associated with a certain 4-fermion condensation occurring well above the electroweak symmetry breaking scale. The point is that a 4-fermion condensate, composed of fermions with standard quantum numbers, can signal a dynamical breaking of isospin symmetry while not breaking $SU(2) \times U(1)$ gauge symmetry. A pattern of symmetry breaking is described [1] in which the isospin breaking feeds strongly into the top mass while not significantly influencing the $W$ and $Z$ masses.

Our other goal is to make a first attempt to extract the momentum dependence of the dynamical fermion 4-point function. For example if such an object is to play a role in generating a fermion mass, then two lines of the 4-point function may be closed off into a loop. The details of the momentum dependence, such as knowing the relative size of the 4-point function when different pairs of momenta are large, is then of interest. As a first guess, it is standard to appeal to some kind of factorization or vacuum saturation and treat the 4-point function as a product of two 2-point functions. We proceed beyond that type of picture here, although we are forced into making our own truncations of the full treatment.

Truncations are also made in the study of the dynamical mass function, but there is growing confidence that most of the essential physics of chiral symmetry breaking is contained in these treatments. For example, the kernel of the SD equation is usually dealt with in some form of ladder approximation. But in the case of a constant coupling, which is the case we study here, it is found that the main conclusions regarding the behavior of the mass function are quite insensitive to the precise form of the kernel [2].

In the present study we use a one gauge-boson exchange approximation. Since the gauge boson can attach to any pair of the four legs, the SD equation sums up a much more complicated set of diagrams than the set of ladder graphs. It is clear from the study of the mass function that the poorly understood infrared effects (e.g. confinement) are not important in extracting the intermediate and high momentum behavior of the mass function. We shall adopt a similar philosophy here, and drop various nonlinearities which arise only in the infrared.

In the next section we consider the SD equation for the fermion 4-point function. In section 3 we present an explicit derivation of the equation within the framework of certain approximations, which are further discussed in section 4. In section 5 we discuss our numerical treatment, and in section 6 we present our results.

## 2  Schwinger-Dyson equation

Our model consists of a strongly interacting U(1) gauge theory with one Dirac fermion. We start by considering 4-fermion operators of the form $\bar\psi_a \Gamma_{ab} \psi_b \bar\psi_c \Gamma'_{cd} \psi_d$, where $\Gamma$ and $\Gamma'$ are $4 \times 4$ matrices and $a, b, c, d$ are spinor indices. In this work, we are particularly interested in operators that are purely chirality changing[1], i.e. of the form $\bar\psi_L \Gamma \psi_R \bar\psi_L \Gamma' \psi_R$ + h.c., where $\psi_L \equiv \frac{(1-\gamma_5)}{2}\psi$ and $\psi_R \equiv \frac{(1+\gamma_5)}{2}\psi$. The two independent operators which have this property and respect parity are[2]

$$\frac{1}{2}(\bar\psi\psi\bar\psi\psi + \bar\psi\gamma^5\psi\bar\psi\gamma^5\psi) = \bar\psi_L\psi_R\bar\psi_L\psi_R + \text{h.c.}$$

$$\bar\psi\sigma^{\mu\nu}\psi\bar\psi\sigma^{\mu\nu}\psi = \bar\psi_L\sigma^{\mu\nu}\psi_R\bar\psi_L\sigma^{\mu\nu}\psi_R + \text{h.c.} \qquad (1)$$

The vacuum expectation values of these two operators receive exclusively non-perturbative contributions, and their non-zero values signal a dynamical break-down of a chiral $U(1)_A$. However the discrete symmetry $e^{i(n/2)\pi\gamma_5}$ where $n$ is an integer can remain unbroken and prevent fermion masses. We imagine that this discrete symmetry is broken on a smaller energy scale in which case the resulting fermion masses can be neglected in our analysis.

We develop the appropriate SD formalism in analogy with the standard analysis of the 2-point function. The SD equation for the 2-point function expresses this function in terms of a 3-point function, as shown diagrammatically in Fig. 1. This equation is part of an infinite SD hierarchy, which must be truncated in some manner. The loop expansion in the effective potential formalism [4] leads to such an approximation, and at two-loops it yields the integral equation given in Fig. 2.

The diagrammatic form of the SD equation for the 4-point function is shown in Fig. 3. It involves a 5-point function, and it is again a part of the infinite SD-equation hierarchy. In the corresponding effective potential we have terms as shown in Fig. 4a, b. The smaller black dots depict the fermion self-energies, and the larger dots the 4-point function. If the fermions have non-zero self-energies, terms of the form shown in Fig. 4a, with and without gluons attached, would be important. The 4-fermion condensate would turn out to be proportional to the square of the 2-fermion condensate, if other diagrams are neglected.

---

[1]For a discussion on chirality-conserving 4-fermion operators, see [3].
[2]We choose to work in Euclidean space, so there is no difference between upper and lower Lorentz indices.

However, if the fermion masses can be neglected, as we assume in this work, then the diagram in Fig. 4b is essential for a nonvanishing 4-point function. Minimization of the effective potential with respect to the 4-point function leads us to an equation of the form shown in Fig. 5. It is clear that the 5-point function of Fig. 3 has been approximated by a 4-point function, with a gauge boson attached to one of the fermion legs with a bare coupling, as shown in Fig. 4c, and in complete analogy with the study of the 2-point function. Note that we have symmetrized the set of diagrams such that a gluon is attached to all possible pairs of fermion legs. We then have to insert a factor of 1/2 to recover the correct SD equation.

The resulting SD equation is linear. Non-linearities would come from vacuum terms like the one shown in Fig. 6a, which contributes to the SD equation a term displayed in Fig. 6b. Note that the number of black dots in Fig. 6a must be even. These non-linear terms are not easily resummable, in contrast to the case of the 2-point function or the Nambu-Jona-Lasinio model. We do not consider these terms in this work since we expect them to affect the 4-point function mostly in the infra-red regime, in analogy with the 2-point function.

We now derive the explicit analytic form of the equation shown diagrammatically in Fig. 5.

# 3  Feynman diagrams

We are considering a 4-point function associated with the Green's function $\left\langle 0|T\{\overline{\psi_a}(x)\psi_b(y)\overline{\psi_c}(z)\psi_d(0)\}|0\right\rangle$. The 4-point function receiving exclusively non-perturbative contributions is, in momentum space

$$\mathcal{O}_{abcd}(s,t,u,p_1{}^2,p_2{}^2,p_3{}^2) = \mathcal{O}_{S+P}(I_{ab} \otimes I_{cd} + \gamma^5_{ab} \otimes \gamma^5_{cd}) + \mathcal{O}_T \sigma^{\mu\nu}_{ab} \otimes \sigma^{\mu\nu}_{cd} \qquad (2)$$

where $p_1, ..., p_4$ are the external momenta corresponding to the fermions with spinor indices $a, ..., d$ respectively. We wish to develop the SD equations for the two scalar functions $\mathcal{O}_{S+P}$ and $\mathcal{O}_T$.

When these functions appear inside the integrals in the SD equation, then their arguments depend on the loop momentum $k$ flowing in the gauge boson propagator. For example, we have the $k$ dependent variables $s^i, t^i$, and $u^i$, where the superscript $i$ is equal to $A, ..., F$, corresponding to each of the six diagrams on the right-hand side of the equation in Fig. 5. They are given as follows:[3]

| | | |
|---|---|---|
| $s = -(p_1 + p_2)^2$ | $t = -(p_1 - p_4)^2$ | $u = -(p_1 - p_3)^2$ |
| $s^A = s$ | $t^A = -(p_1 - p_4 - k)^2$ | $u^A = -(p_1 - p_3 - k)^2$ |
| $s^B = s$ | $t^B = -(p_1 - p_4 + k)^2$ | $u^B = u^A$ |
| $s^C = -(p_1 + p_2 - k)^2$ | $t^C = t$ | $u^C = u^A$ |
| $s^D = -(p_1 + p_2 + k)^2$ | $t^D = t$ | $u^D = u^A$ |
| $s^E = s^C$ | $t^E = t^A$ | $u^E = u$ |
| $s^F = s^D$ | $t^F = t^A$ | $u^F = u,$ |

The 4-point functions with arguments as they would appear inside the six diagrams are denoted by $\mathcal{O}^i_{S+P}$ and $\mathcal{O}^i_T$, where $i = A, ..., F$.

---

[3] We define these variables with a minus sign because we work in Euclidean space, and we want to work with positive quantities.

We then define the following functional operators $\Gamma^i$:

$$\begin{aligned}
\Gamma^A[K] &\equiv \frac{\alpha}{8\pi^3} \int d^4k \frac{K}{(p_1-k)^2(p_2+k)^2} \\
\Gamma^B[K] &\equiv \frac{\alpha}{8\pi^3} \int d^4k \frac{K}{(p_3+k)^2(p_4-k)^2} \\
\Gamma^C[K] &\equiv \frac{\alpha}{8\pi^3} \int d^4k \frac{K}{(p_1-k)^2(p_4-k)^2} \\
\Gamma^D[K] &\equiv \frac{\alpha}{8\pi^3} \int d^4k \frac{K}{(p_2+k)^2(p_3+k)^2} \\
\Gamma^E[K] &\equiv \frac{\alpha}{8\pi^3} \int d^4k \frac{K}{(p_1-k)^2(p_3-k)^2} \\
\Gamma^F[K] &\equiv \frac{\alpha}{8\pi^3} \int d^4k \frac{K}{(p_2+k)^2(p_4+k)^2},
\end{aligned} \quad (3)$$

where $K$ is a function of the loop and external momenta, with a possibly non-trivial spinor structure, and $\alpha$ is the momentum-independent coupling. We work in the Landau gauge which is popular in studies of SD equations; the gauge boson propagator reads $\frac{D^{\mu\nu}}{k^2} \equiv \frac{1}{k^2}\left(\delta^{\mu\nu} - \frac{k^\mu k^\nu}{k^2}\right)$.

With regards to the gauge dependence of our results we note that similar gauge dependence occurs in the two-point SD equation, and that the choice of Landau gauge in that case gives a critical coupling resembling that found in more complete treatments [5], [6]. Since one of our main objectives is to compare the critical coupling for the two-point and four-point functions, it seems natural to use the same approximation (ladder) and the same gauge (Landau). We also note that in our case of massless fermions the one-loop fermion self-energy vanishes in Landau gauge, thus making the use of the bare photon-fermion vertex consistent with the Ward-Takahashi identity. To do better and to approach our problem in a truly gauge invariant way, we would have to carry over the sort of treatment presented in [5]; this is a task far beyond our present means.

To significantly reduce the complexity of our problem we make the following truncation. Inside the loop integrals we drop certain terms which are suppressed in the large loop-momentum limit. In particular, we drop terms proportional to $\not{p}_i$, $i = 1...4$, coming from the fermion propagators, and we replace $k^\mu k^\nu$ terms, where $k$ is the loop momentum, by $\frac{k^2}{4}\delta^{\mu\nu}$. An example of the effect of this type of truncation in a SD equation can be found in the literature. The authors of Ref.[7] used a 3-point SD equation for a calculation of the pion decay constant $f_\pi$. They found that results obtained by neglecting terms in the pseudoscalar vertex proportional to external momenta were surprisingly very similar to their full results. The truncation in that case turns out to correspond to the popular Pagels-Stokar approximation [8] for the calculation of $f_\pi$. Our case is of course quite different, and we will test further the effect of such a truncation below.

In the following, by "Scalar insertion" we mean placing the function $\mathcal{O}_S \cdot I \otimes I$ inside the integral on the right-hand side of the SD equation, by "Pseudoscalar insertion" placing $\mathcal{O}_P \cdot \gamma^5 \otimes \gamma^5$, and by "Tensor insertion" placing $\mathcal{O}_T \cdot \sigma^{\mu\nu} \otimes \sigma^{\mu\nu}$. The spinor structure can be simplified as follows:

I. Diagrams A and B

a) Scalar insertion:
$$\Gamma^{A,B} \left[ \frac{\mathcal{O}_S^{A,B}}{k^2} D^{\mu\nu} \gamma^\nu \not{k}\not{k}\gamma^\mu \right] \otimes I = 3\Gamma^{A,B}[\mathcal{O}_S^{A,B}] I \otimes I$$

b) Pseudoscalar insertion:
$$\Gamma^{A,B} \left[ \frac{\mathcal{O}_P^{A,B}}{k^2} D^{\mu\nu} \gamma^\nu \not{k}\gamma^5 \not{k}\gamma^\mu \right] \otimes \gamma^5 = 3\Gamma^{A,B}[\mathcal{O}_P^{A,B}]\gamma^5 \otimes \gamma^5$$

c) Tensor insertion:
$$\Gamma^{A,B} \left[ \frac{\mathcal{O}_T^{A,B}}{k^2} D^{\mu\nu} \gamma^\nu \not{k}\sigma^{\tau\rho}\not{k}\gamma^\mu \right] \otimes \sigma^{\tau\rho} = -\Gamma^{A,B}[\mathcal{O}_T^{A,B}]\sigma^{\mu\nu} \otimes \sigma^{\mu\nu}$$

II. Diagrams C and D

a) Scalar insertion:
$$\Gamma^{C,D} \left[ \left( \frac{\mathcal{O}_S^{C,D}}{k^2} D^{\mu\nu} \gamma^\nu \not{k} \right) \otimes (\not{k}\gamma^\mu) \right] = \tfrac{1}{4}\Gamma^{C,D}[\mathcal{O}_S^{C,D}]\sigma^{\mu\nu} \otimes \sigma^{\mu\nu}$$

b) Pseudoscalar insertion:

$$\Gamma^{C,D}\left[\left(\frac{\mathcal{O}_P^{C,D}}{k^2}D^{\mu\nu}\gamma^\nu \slashed{k}\gamma^5\right) \otimes (\gamma^5 \slashed{k}\gamma^\mu)\right] = \tfrac{1}{4}\Gamma^{C,D}[\mathcal{O}_P^{C,D}]\sigma^{\mu\nu} \otimes \sigma^{\mu\nu}$$

c) Tensor insertion:

$$\Gamma^{C,D}\left[\left(\frac{\mathcal{O}_T^{C,D}}{k^2}D^{\mu\nu}\gamma^\nu \slashed{k}\sigma^{\tau\rho}\right) \otimes (\sigma^{\tau\rho}\slashed{k}\gamma^\mu)\right] =$$
$$2\Gamma^{C,D}[\mathcal{O}_T^{C,D}]\sigma^{\mu\nu} \otimes \sigma^{\mu\nu} + 6\Gamma^{C,D}[\mathcal{O}_T^{C,D}](I \otimes I + \gamma^5 \otimes \gamma^5)$$

where we used the identity

$$\gamma^\alpha \gamma^\beta \gamma^\lambda = -\delta^{\alpha\beta}\gamma^\lambda - \delta^{\beta\lambda}\gamma^\alpha + \delta^{\alpha\lambda}\gamma^\beta + i\gamma^5 \varepsilon^{\alpha\beta\lambda\mu}\gamma^\mu.$$

III. Diagrams E and F

a) Scalar insertion:
$$\Gamma^{E,F}\left[\left(\frac{\mathcal{O}_S^{E,F}}{k^2}D^{\mu\nu}\gamma^\nu \slashed{k}\right) \otimes (\gamma^\mu \slashed{k})\right] = -\tfrac{1}{4}\Gamma^{E,F}[\mathcal{O}_S^{E,F}]\sigma^{\mu\nu} \otimes \sigma^{\mu\nu}$$

b) Pseudoscalar insertion:
$$\Gamma^{E,F}\left[\left(\frac{\mathcal{O}_P^{E,F}}{k^2}D^{\mu\nu}\gamma^\nu \slashed{k}\gamma^5\right) \otimes (\gamma^\mu \slashed{k}\gamma^5)\right] = -\tfrac{1}{4}\Gamma^{E,F}[\mathcal{O}_P^{E,F}]\sigma^{\mu\nu} \otimes \sigma^{\mu\nu}$$

c) Tensor insertion:
$$\Gamma^{E,F}\left[\left(\frac{\mathcal{O}_T^{E,F}}{k^2}D^{\mu\nu}\gamma^\nu \slashed{k}\sigma^{\tau\rho}\right) \otimes (\gamma^\mu \slashed{k}\sigma^{\tau\rho})\right] =$$
$$2\Gamma^{E,F}[\mathcal{O}_T^{E,F}]\sigma^{\mu\nu} \otimes \sigma^{\mu\nu} - 6\Gamma^{E,F}[\mathcal{O}_T^{E,F}](I \otimes I + \gamma^5 \otimes \gamma^5).$$

We may now equate terms proportional to $I \otimes I$, $\gamma^5 \otimes \gamma^5$ and $\sigma^{\mu\nu} \otimes \sigma^{\mu\nu}$ on either side of SD equation, in order to obtain the following system of coupled integral equations:

$$\begin{aligned}
\mathcal{O}_{S+P} &= 3(\overline{\Gamma^A} + \overline{\Gamma^B})[\mathcal{O}_{S+P}] + \\
&\quad 6(\overline{\Gamma^C} + \overline{\Gamma^D} - \overline{\Gamma^E} - \overline{\Gamma^F})[\mathcal{O}_T] \\
\mathcal{O}_T &= \left(-(\overline{\Gamma^A} + \overline{\Gamma^B}) + 2(\overline{\Gamma^C} + \overline{\Gamma^D} + \overline{\Gamma^E} + \overline{\Gamma^F})\right)[\mathcal{O}_T] \\
&\quad + \frac{1}{2}(\overline{\Gamma^C} + \overline{\Gamma^D} - \overline{\Gamma^E} - \overline{\Gamma^F})[\mathcal{O}_{S+P}],
\end{aligned} \quad (4)$$

where we have defined $\overline{\Gamma^i}[\mathcal{O}] \equiv \Gamma^i[\mathcal{O}^i], i = A, ..., F$.

# 4 Further simplification

This is still a formidable system of integral equations when it is realized that the quantities $\mathcal{O}_{S+P}$ and $\mathcal{O}_T$ are functions of six variables, $s$, $t$, $u$, $p_1^2$, $p_2^2$, and $p_3^2$. The most interesting dependence is on the three variables $s$, $t$, and $u$, and we would be happy if we were able to solve for the function

$$\hat{\mathcal{O}}(s,t,u) \equiv \mathcal{O}(s,t,u,p^2,p^2,p^2) \tag{5}$$

where $4p^2 = s + t + u$. But the SD equation does not permit this, since for loop momenta much larger than the external momenta we must know the value of $\mathcal{O}$ with two of the $p_i$'s much larger than the other two.

In order to proceed we shall make the following replacement inside the integrals:

$$\mathcal{O}(s',t',u',p_1'^2,p_2'^2,p_3'^2) \to \hat{\mathcal{O}}(s',t',u')\sqrt{\frac{|\hat{\mathcal{O}}(s,t,u)|}{|\hat{\mathcal{O}}(s',t',u')|}}. \tag{6}$$

A prime indicates a possible dependence on the loop momentum $k$. For large loop momenta, we find it reasonable to assume that the value of $\mathcal{O}$ lies between its value when the large momentum flows through all of its legs and its value when the large momentum is removed. We have chosen the geometric mean.

This prescription yields equations which can now be solved for the function $\hat{\mathcal{O}}(s,t,u)$. The final equations are

$$\begin{aligned}
O_{S+P} &= 3(\overline{\Gamma^A} + \overline{\Gamma^B})[O_{S+P}] + \\
&\quad \frac{6|O_T|}{|O_{S+P}|}(\overline{\Gamma^C} + \overline{\Gamma^D} - \overline{\Gamma^E} - \overline{\Gamma^F})[O_T] \\
O_T &= \left(-(\overline{\Gamma^A} + \overline{\Gamma^B}) + 2(\overline{\Gamma^C} + \overline{\Gamma^D} + \overline{\Gamma^E} + \overline{\Gamma^F})\right)[O_T] \\
&\quad + \frac{|O_{S+P}|}{2|O_T|}(\overline{\Gamma^C} + \overline{\Gamma^D} - \overline{\Gamma^E} - \overline{\Gamma^F})[O_{S+P}],
\end{aligned} \tag{7}$$

where the function $O$ is defined by the relation $\hat{\mathcal{O}} \equiv |O|O$.

We can provide more motivation for the prescription in (6) as follows. If we were to drop the four terms depending on $O_T$ in the equation for $O_{S+P}$, we would have a much simpler equation in which it is possible to set $s = 0$ and in which the solution depends on $t+u$ only. In fact, if we make the replacements $O_{S+P} \to \Sigma$ and $(t+u) \to p^2$ then we would be left with the linearized SD equation for the two-point function used to study the dynamical mass $\Sigma(p^2)$. The critical coupling would be the same, since we have the two diagrams $A$ and $B$ in Fig.5 which give identical one-gauge-boson exchange contributions in this case, and a factor of $1/2$.

In our case of constant coupling this equation is well-known to possess a solution of the form $\Sigma(p^2) \approx 1/p$ up to logs. We therefore deduce that the result for $\hat{\mathcal{O}}_{S+P} \equiv |O_{S+P}|O_{S+P}$ is such that it falls like $1/(t+u)$. This agrees with the fact that we are here summing the set of ladder graphs generated by diagrams $A$ and $B$, and therefore the solution should behave like $\Sigma(p^2)^2 \approx 1/p^2$ when large momentum flows though both halves of the diagram and $s = 0$. We see that this consistency is a direct result of the prescription in (6), and we therefore proceed to use this prescription in the full set of equations. Note that the above discussion provides a naive guess as to the size of the critical coupling; it will be of interest to see how this survives in the full set of equations.

## 5  Numerical Analysis

We discretize the three-dimensional argument space spanned by $s$, $t$, and $u$, and the four-dimensional integration momentum space. We take the integration 4-momentum to have the form $k_\mu = (k_0, |k|\sin\tilde{\theta}\cos\tilde{\phi}, |k|\sin\tilde{\theta}\sin\tilde{\phi}, |k|\cos\tilde{\theta})$. Each of the variables $s$, $t$, $u$, $|k|^2$ and $k_0^2$ are spaced logarithmically according to $\log_{10}(\Lambda_{IR}^2) + \frac{i}{n}\log_{10}(\Lambda_{UV}^2/\Lambda_{IR}^2)$, where $i = 1, ..., n$. We discretize the integration angles by the prescriptions $\phi(i) = 2\pi i/n$ and $\cos\tilde{\theta}(i) = -1 + 2i/(n+1)$. We do not take the points $\tilde{\theta} = 0, \pi$ as integration points, because the kernel has an integrable (collinear) singularity there. The ratio of the ultraviolet and infrared cutoffs, $\Lambda_{UV}/\Lambda_{IR}$, should not be too small, but we cannot take it too large either because of the limited number of points on the lattice.

Since we have specialized to $p_i^2 = p^2$ for all $i = 1, ..., 4$, we may take a reference frame in which

$$
\begin{aligned}
p_1 &= (p_0, -|p|, 0, 0) \\
p_2 &= (p_0, |p|, 0, 0) \\
p_3 &= (p_0, |p|\cos\theta, |p|\sin\theta, 0) \\
p_4 &= (p_0, -|p|\cos\theta, -|p|\sin\theta, 0)
\end{aligned}
\quad (8)
$$

In this frame, $s = 4p_0^2$, $u = 4|p|^2(1 + \cos\theta)$ and $t = 4|p|^2(1 - \cos\theta)$. For each point on the $s$, $t$, $u$ lattice we calculate $p_0$, $|p|$, and $\theta$, and then use these values to determine the kernels of the integral equations and the $k$-dependent arguments $s^i, t^i, u^i$ ($i = A, ..., F$) of the 4-point functions. Note that the 4-point functions are only defined for certain values of their arguments $s$, $t$, $u$ according to the initial discretization. However, these 4-point functions are needed at different values of their arguments inside the integrals, due to the loop momentum. The values of the 4-point functions at these points are found by linear interpolation.

The relaxation method used consists of inserting initial configurations for the 4-point functions $O_{S+P}$ and $O_T$, and then iterating (4) until it is satisfied to a reasonable accuracy. For the results we present, the number of points on which the 4-point functions are defined is $12^3$ and the number of integration points is $12^4$. Moreover, we choose $\Lambda_{UV}/\Lambda_{IR} = 6$. The general features of the 4-point functions do not vary in any drastic way either by increasing the total number of points from $6^7$ to $12^7$, or by varying the $\Lambda_{UV}/\Lambda_{IR}$ ratio.

# 6 The results

Our equations exhibit critical behavior in the sense that the coupling must be sufficiently large to support a nonvanishing solution. For $\Lambda_{UV}/\Lambda_{IR} = 6$ we find a critical coupling

$$\frac{\alpha_c}{\alpha_c^{2\text{pt}}} \approx 0.96 \pm 0.1. \tag{9}$$

We denote by $\alpha_c^{2\text{pt}}$ the critical coupling required for a chiral symmetry breaking solution in the 2-point function. To be consistent, we calculate $\alpha_c^{2\text{pt}}$ using our 4-point function algorithm by keeping only diagrams $A$ and $B$, setting $s = 0$, and using the same $\Lambda_{UV}/\Lambda_{IR}$. By our previous discussion this is equivalent (up to our approximations) to the 2-point SD equation. The ratio $\frac{\alpha_c}{\alpha_c^{2\text{pt}}}$ is not considerably affected by varying the $\Lambda_{UV}/\Lambda_{IR}$ ratio. Our estimated error in $\alpha_c$ reflects only the effect of the finite number of points used, which limits the accuracy of the integration and the interpolation procedure used inside the integrals. Rather surprisingly, it appears that our naive guess above as to the size of the critical coupling survives in the full equations.

The numerical solution of (4) is shown in Figs. 7-12. In Figs. 7-10 we give the functional dependence of the function $\hat{\mathcal{O}}_{S+P} \equiv |O_{S+P}|O_{S+P}$ on $\tilde{t} \equiv t/\Lambda_{IR}^2$ and on $\tilde{u} \equiv u/\Lambda_{IR}^2$, for two different values of $\tilde{s} \equiv s/\Lambda_{IR}^2$. We use a logarithmic scale and only show points which are, in absolute value, greater than $10^{-4}$. Note that for $u < t$, which in our frame of reference corresponds to angles $\theta > \pi/2$, the 4-point function is negative. In that region we plot the function $-\hat{\mathcal{O}}_{S+P}$, as shown in Figs. 8 and 10.[4] In Figs. 11-12 we give the analogous results for the function $\hat{\mathcal{O}}_T \equiv |O_T|O_T$. This function is always positive, although it comes close to vanishing for large $t$ and $u$.

---

[4]Note that the perspective on these figures is rotated by $\pi$ with respect to Figs. 7 and 9.

One of the main features of the 4-point functions is their slow fall with $s$. Another is that $\hat{\mathcal{O}}_{S+P}$ is roughly antisymmetric and $\hat{\mathcal{O}}_T$ roughly symmetric with respect to the interchange $t \leftrightarrow u$. These symmetry properties are quite consistent with the form of the full integral equations in (7). When the solutions are inserted into these equations we see that all terms in the $\hat{\mathcal{O}}_{S+P}$ equation are antisymmetric under the interchange $t \leftrightarrow u$, while all terms in the $\hat{\mathcal{O}}_T$ equation are symmetric. This follows from the way the $C$ and $D$ diagrams are related to the $E$ and $F$ diagrams when $t \leftrightarrow u$. Note that a solution with $\hat{\mathcal{O}}_{S+P}$ symmetric and $\hat{\mathcal{O}}_T$ antisymmetric under $t \leftrightarrow u$ would also be consistent, but it is disfavored by the equations.

The order of magnitude of $|O_{S+P}|$ and $|O_T|$ is the same when the angle $\theta$ is not too close to $\pi/2$. This means that the ratios of the 4-point functions multiplying the coupling terms in (7) do not have a major effect; note that the appearance of these ratios is the main change from the equations in (4). In certain regions in the plots the 4-point functions display an almost linear slope (in logarithmic coordinates). Both $|\hat{\mathcal{O}}_{S+P}|$ and $\hat{\mathcal{O}}_T$ fall approximately as $1/t^3$ while remaining almost constant with $u$ when $t < u$, and fall approximately as $1/u^3$ while remaining almost constant with $t$ when $t > u$. The plots also display some structure (ridges etc.) on a smaller scale. We believe that these latter features are artifacts of our various truncations and/or the discretization.

We would like some idea as to the effect of our procedure of dropping terms proportional to the external momenta. One check is obtained by solving the system of integral equations when the terms coupling the two equations with each other are dropped. We plot the corresponding result for $\hat{\mathcal{O}}_{S+P}$ in Fig. 13, and we find a result which is falling roughly as $1/(t+u)$ except for the highest values of $t+u$. This behavior over most $t + u$ is consistent with our discussion above showing the correspondence of this case to the two-point function. The omission of terms in the integral proportional to the external momentum causes a distortion in the solution only at the high end of $t + u$.

By comparing Fig. 13 with our full result for $\hat{\mathcal{O}}_{S+P}$, we can see that the terms which couple the two full equations together have a very large effect on the form of the solution. Most dramatic is the very strong dependence of $\hat{\mathcal{O}}_{S+P}$ on $t - u$. In the case of $\hat{\mathcal{O}}_T$ there is less difference between full result and the decoupled equation result displayed in Fig. 14. Also note that in the decoupled equation for $\hat{\mathcal{O}}_{S+P}$, the variable $s$ enters in a simpler way and we are able to set it equal to a constant (the infrared cut-off in Fig. 13). This is not possible in the full equations. Nevertheless, we find a fairly weak dependence on $s$ in the full results.

We have also made the following tests. When external momentum terms are dropped, we are distorting the integrand for loop momenta of order than or less than the external momenta. We may thus try making other distortions of the full equations in the infrared loop-momentum region and see the consequences. As one distortion, we apply an infrared cutoff on the loop momentum at a $k^2$ equal to the $p^2$ in (5). As another distortion, we replace all appearances of $(P-k)^2$, where $P$ is one or some combination of the external momenta, by the quantity $\max(P^2, k^2)$, which is angle independent. In both cases, we find that the resulting 4-point functions are similar to those presented above, except for some of the highest values of $s$, $t$, and $u$.

## 7 Conclusions

In this work, we have treated a linearized and truncated Schwinger-Dyson equation for the 4-point function of fermions interacting via a strongly interacting U(1) gauge interaction in the Landau gauge. We considered chirality-changing 4-point functions which receive nonperturbative contributions exclusively. Our basic philosophy in dealing with the full intractable problem is to make various truncations to it which allow solutions to be found, and then identify those features of the solutions which are generic. This work represents a first step in this direction.

Thus far, our numerical results indicate that the equations exhibit a critical behavior, and that the critical coupling appears to be roughly equal to the one required for the formation of 2-fermion condensates. Our results also provide a first indication of the momentum dependence of the 4-point functions. Further studies and possible ramifications will be discussed elsewhere.

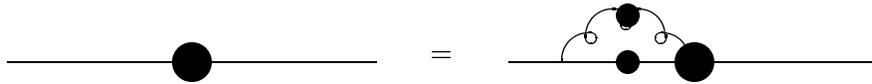

Figure 1: The schematic form of the SD equation for the 2-point function.

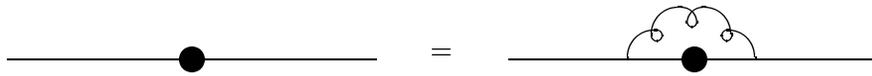

Figure 2: The schematic form of the SD equation in ladder approximation.

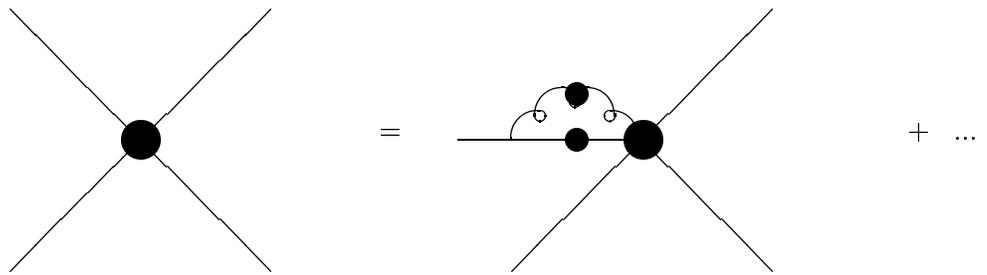

Figure 3: The schematic form of the SD equation for the 4-point function.

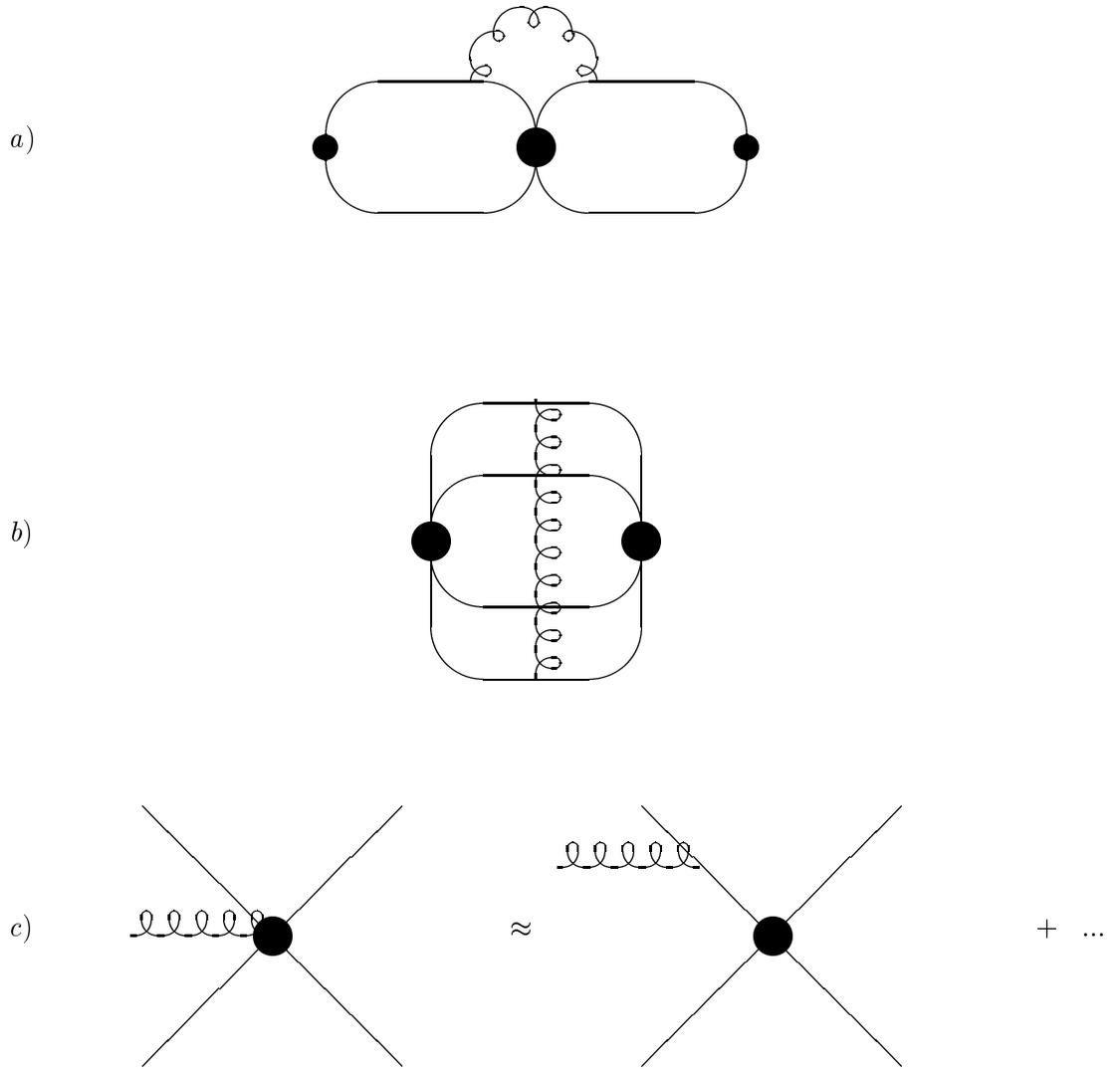

Figure 4: a) A term in the effective potential. b) An example of a leading term at first order in $\alpha$, when the fermions are massless. c) The implied approximation.

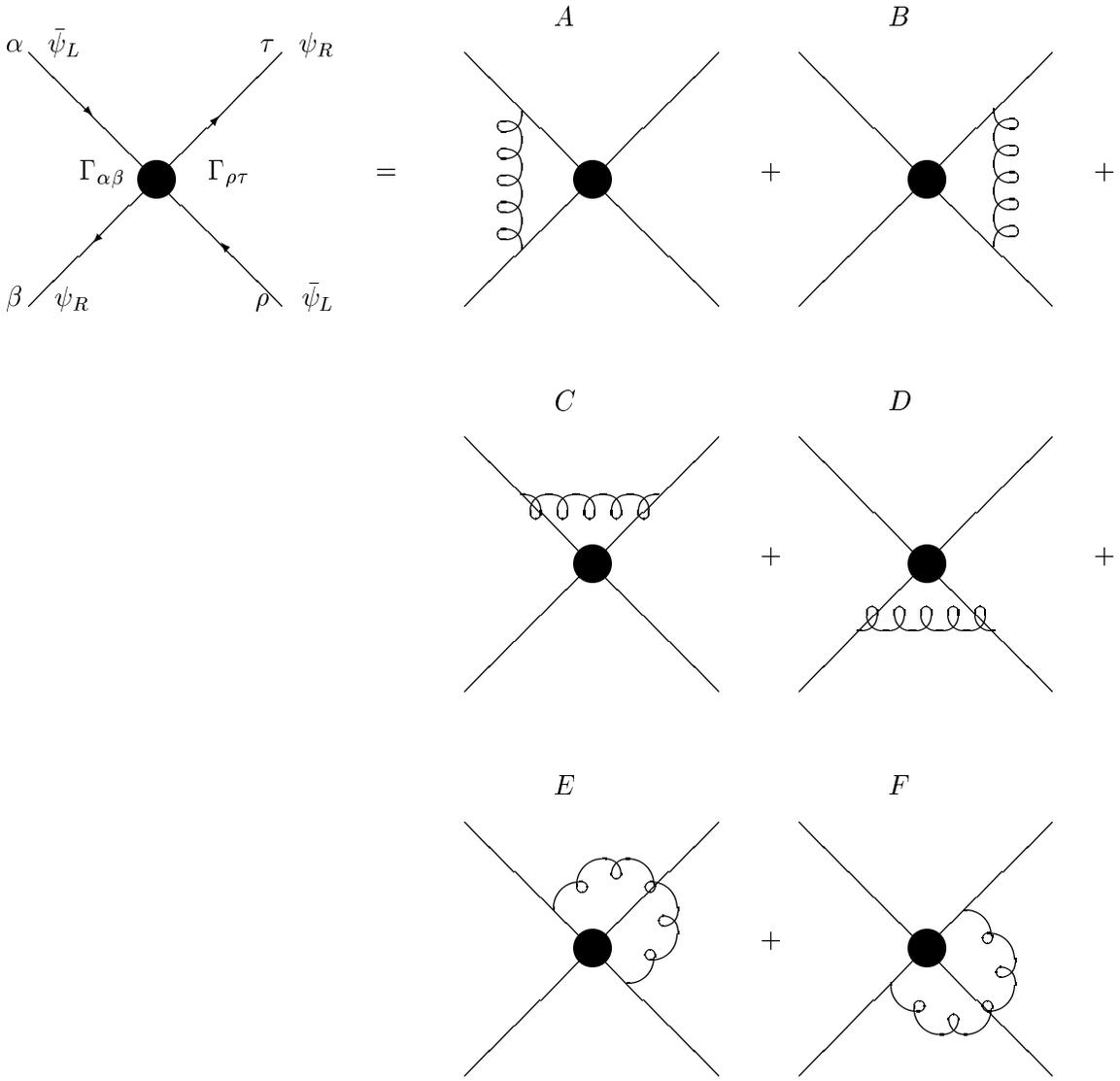

Figure 5: The schematic form of the SD equation. We have labeled the four fermions by their spinor indices. We label the diagrams in the right-hand side by the capital letters $A, ..., F$. We omit the factor of $\frac{1}{2}$ multiplying the right-hand side.

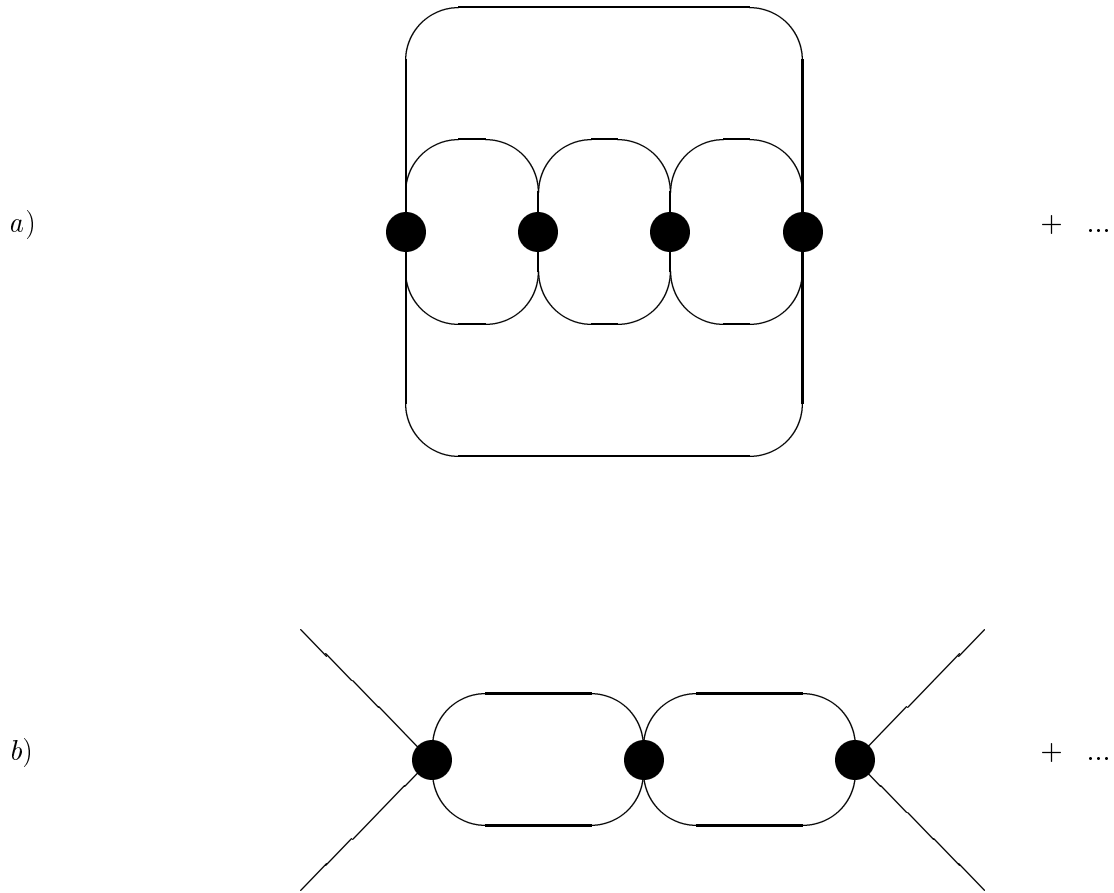

Figure 6: a) The first of the vacuum terms in the effective potential which would lead to a non-linear SD equation. b) The corresponding contribution to the SD equation.

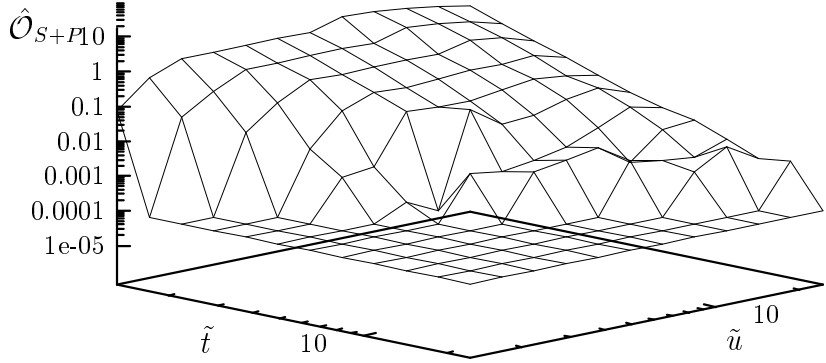

Figure 7: The 4-point function $\hat{\mathcal{O}}_{S+P}$ when $\hat{\mathcal{O}}_{S+P} > 0$, for $\tilde{s} = 1$.

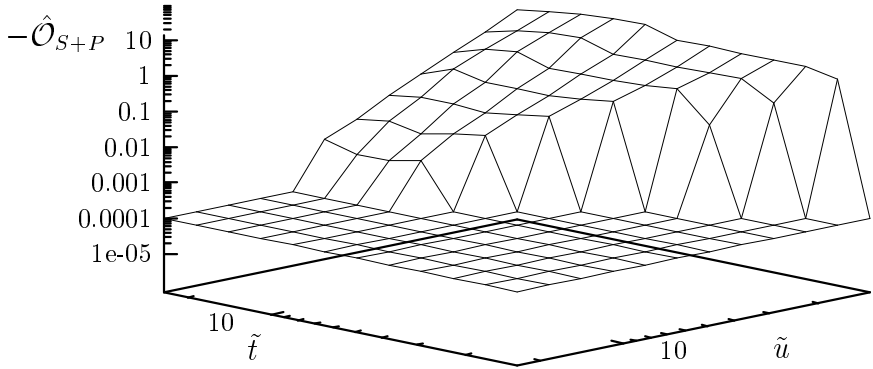

Figure 8: The 4-point function $-\hat{\mathcal{O}}_{S+P}$ when $\hat{\mathcal{O}}_{S+P} < 0$, for $\tilde{s} = 1$.

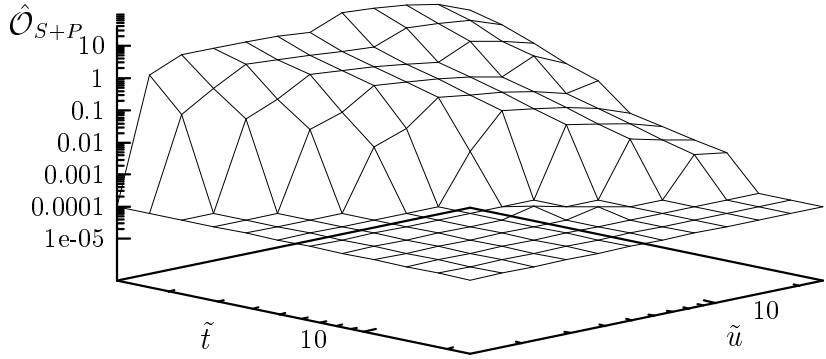

Figure 9: The 4-point function $\hat{\mathcal{O}}_{S+P}$ when $\hat{\mathcal{O}}_{S+P} > 0$, for $\tilde{s} = 25$.

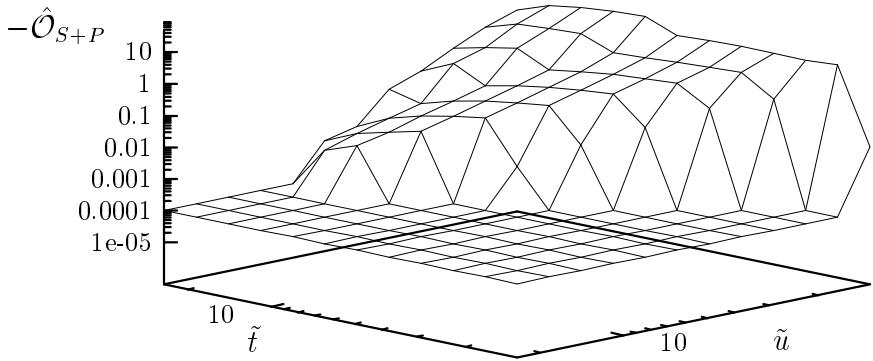

Figure 10: The 4-point function $-\hat{\mathcal{O}}_{S+P}$ when $\hat{\mathcal{O}}_{S+P} < 0$, for $\tilde{s} = 25$.

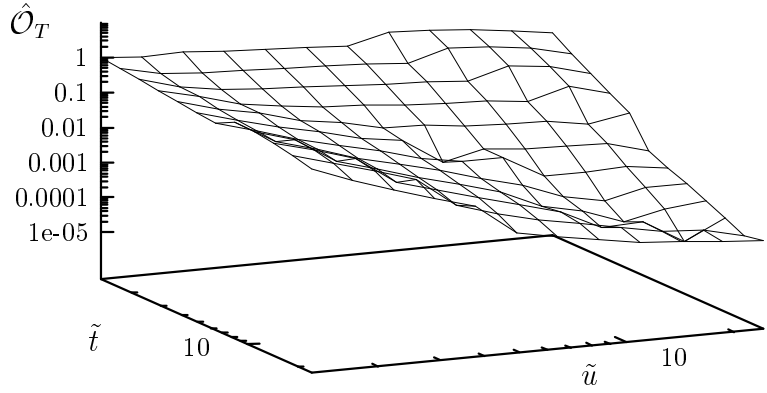

Figure 11: The 4-point function $\hat{\mathcal{O}}_T$ for $\tilde{s} = 1$.

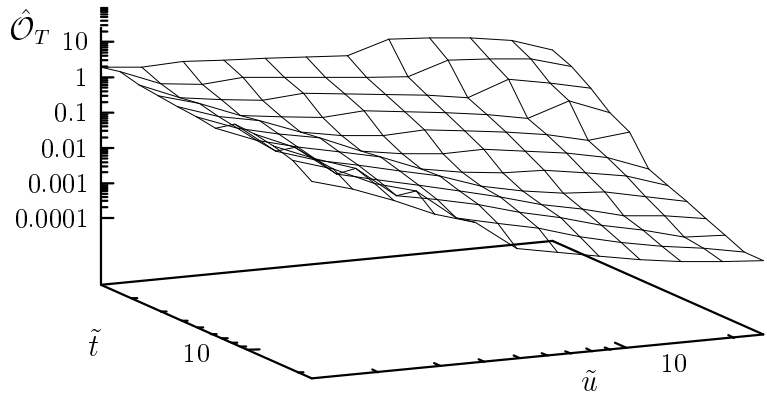

Figure 12: The 4-point function $\hat{\mathcal{O}}_T$ for $\tilde{s} = 25$.

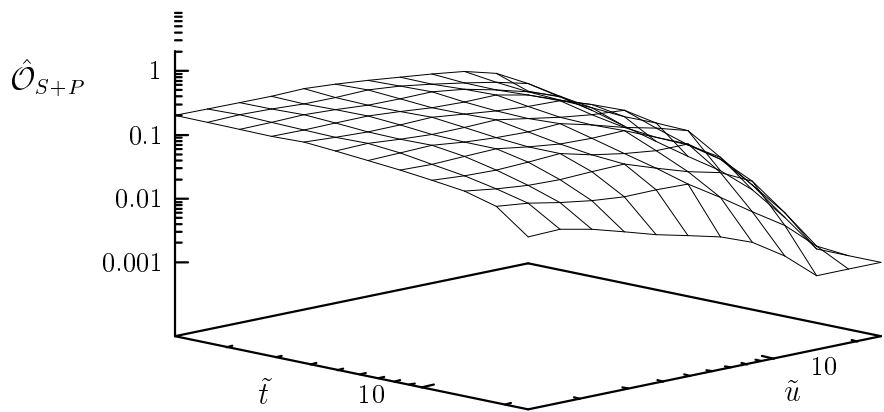

Figure 13: The 4-point function $\hat{\mathcal{O}}_{S+P}$ for $\tilde{s} = 1$, when the terms coupling the two equations are dropped.

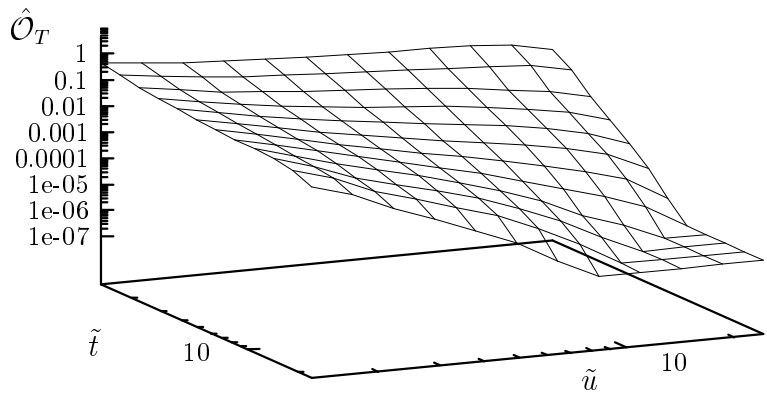

Figure 14: The 4-point function $\hat{\mathcal{O}}_T$ for $\tilde{s} = 1$, when the terms coupling the two equations are dropped.